\documentclass[
 reprint,
showpacs,
showkeys,
amsmath,
amssymb,
aps,
prb,
]{revtex4-1}

\usepackage{graphicx}
\usepackage{dcolumn}
\usepackage{bm}
\usepackage[all]{xy}

\newcommand{\sta}[1]{|#1\!\!>}
\newcommand{\stacc}[1]{<\!\!#1}

\begin{document}

\preprint{APS/123-QED}

\title{Relativistic locality and the action reaction principle\\
predict de Broglie fields} 

\author{Carlos L\'opez}

\affiliation{
Department of Mathematics and Physics, UAH,
28873 Alcal\'a de Henares (Madrid) SPAIN,
carlos.lopez@uah.es}

\date{\today} 

\begin{abstract}
The action reaction principle is violated in the standard formulation of Quantum Mechanics,
so that its phase (Hilbert) space is incomplete. Moreover, projection of state of a quantum system under indirect measurement, when there are alternative virtual paths and one of them is discarded by negative detection implies, according to the action reaction principle, a reaction on the detector, although its macroscopic state does not change. If all interactions are local, mediated by fields with  relativistically causal evolution,
some system different from the particle (which follows another path) must locally interact with the detector. Relativistic locality and the action reaction principle
predict the existence of de Broglie fields.
It is presented a formulation of Quantum Mechanics in extended Hilbert spaces, where 
kinematic and dynamical representations of physical magnitudes are distinguished.

\end{abstract}


\maketitle



\section{\label{sec1}Incompleteness of Quantum Mechanics}

The action reaction principle\cite{newton} (ARP)  
is a fundamental ingredient in the description of interactions, either me\-dia\-ted by local fields or  acting at a distance. 

\hskip 0.2truecm

\centerline{ARP:{\it When two systems interact}}
\centerline{{\it both of them change of state.}}

\hskip 0.2truecm

Relativistic locality\cite{sr} (RL) or causality is grounded in the equivalence of reference frames (independence of the observer), constancy of the light speed
and the arrow of time, distinction between past and future light cones.
The time orientation of a space like curve is frame dependent and can not represent the time evolution of a corpuscular system. The value of a distributed field at a space time event is independent of (or commutes with)  its values at spatially separated events; it is completely determined by initial conditions along a spatial sheet inside the past light cone.

\hskip 0.2truecm

\centerline{RL:{\it The evolution of particles and fields}}
\centerline{{\it fulfils relativistic causality.}}

\hskip 0.2truecm

All known interactions are mediated by local fields. Local action of a field over a particle is accompanied by a local reaction over the field that obeys a causal propagation. Both principles, justified by general considerations of symmetry, have overwhelming empirical support. 

\subsection{The ARP in Quantum Mechanics} 

Let us consider the following interaction in Quantum Mechanics (QM). Two systems 
$\mathcal {S}_I$ and $\mathcal {S}_{II}$ are in initial states  $\sta{a}_I$ and 
$\sta{c}_{II}$, eigenstates of magnitudes $A$ and $C$ respectively. 
$\sta{c}_{II} =$ $\sum _jz_j\sta{d_j}_{II}$ in a basis of eigenstates of a non commuting magnitude $D$. A Hamiltonian of interaction 
$H(A,D) = \kappa  AD$ is switched on. The time evolution of the composite is

\[
\sta{a}_I\!\otimes \sum _jz_je^{-i\kappa a d_jt/\hbar}\sta{d_j}_{II}.
\]
Similarly, there is trivial projection of state after measurement of magnitude $A$ on system $\mathcal {S}_I$ with initial eigenstate $\sta{a}_I$, output $a$,  while the apparatus pointer's state changes. These are academic examples, but the point is that the ARP is {\bf not automatically} fulfilled in the standard mathematical formulation of QM. Its phase space is incomplete, in agreement with the celebrated EPR paper\cite{EPR}.  Additional degrees of freedom in an extended phase space for system $\mathcal {S}_I$ must evolve describing the reaction. 
A mathematical formulation of QM in an extended phase space will be compatible with standard QM if the assignment of probabilities to states in the new formalism reproduces the  probabilities for the corresponding projected states in usual phase space of QM. There are consistent formulations of QM in extended Hilbert spaces\cite{torres}\cite{mio}\cite{mio2} (EQM). The mathematical machinery is preserved: linear superposition in a Hilbert space, operators fulfilling canonical commutation rules, uncertainty principle, Born's rule\cite{Born},\ldots 
But roles of the labels, coordinates of state, and of the dynamical operators associated to physical magnitudes are distinguished, contrarily to orthodox QM. Elementary vectors of state are characterised by joint values of commuting and non commuting magnitudes, 
and play a kinematic role, i.e. description of the evolution of state's magnitudes and  (observable and unobservable) distributions of probability. On the other hand, the self adjoint operators representing magnitudes and  fulfilling the commutation rules play a dynamical role through the Hamiltonian.

\subsection{Relativistic locality and divergent virtual paths}

Particles can follow two or more spatially divergent {\it virtual} paths. An initial state 
$\sta{\Psi}(0)$ splits, e.g. through a Stern--Gerlach apparatus or a beam splitter, and evolves in time to the state

\[
\sta{\Psi}(t_{int})=U(t_{int},0)\sta{\Psi}(0)
=z_1\sta{x_1}+z_2\sta{x_2}  
\] 
where $x_j$ is the spatial position at $t_{int}$ of the  wave packet following
the $j$--th trajectory, and $z_j\sta{x_j}$ is the associated state's component; other quantum numbers are omitted. An appropriate experimental set up can rejoin these paths around $x_f$ at $t_f$, for a final state 
$\sta{\Psi}(t_f)=U(t_f,t_{int})\sta{\Psi}(t_{int})$ where both components interfere. If in another run a particle detector $D^{(1)}$ is placed at $x_1$ either the particle is detected (and blocked) or it arrives to $x_f$ following path $2$.
The state of the composite is 
$\sta{\Psi}(t_{int})\otimes\sta{D^{(1)}_0}$ right before the system/detector interaction, 
$\sta{D^{(1)}_0}$ ready--to--measure state of the pointer. After interaction the composite has evolved to the entangled state  

\[
z_1\sta{x_1}\!\otimes\sta{D^{(1)}_p}+z_2\sta{x_2}\!\otimes\sta{D^{(1)}_n}\, ,
\]
$\sta{D^{(1)}_p}$ ($\sta{D^{(1)}_n}$)  state of the pointer
for positive (negative) detection.
In case of  negative detection and applying the projection of state rule we obtain
$\sta{\Psi_n}(t_{int})=$ $\sta{x_2}\!\otimes\sta{D^{(1)}_n}$, with the particle localised around $x_2$. Next, the particle evolves to the  final state  
$\sta{\Psi_n}(t_f)=$ $U(t_f,t_{int})\sta{x_2}$, localised around $x_f$. 
$\sta{\Psi_n}(t_{int})$ $\neq \sta{\Psi}(t_{int})$, as well as  $\sta{\Psi_n}(t_f)$  
$\neq\sta{\Psi}(t_f)$, so that
$\sta{D^{(1)}_0}$ $\neq \sta{D^{(1)}_n}$ according to the ARP. But when using a particle detector its macroscopic state does not change. Appropriately designed devices could show an observable reaction.

This is a generic experimental set up for a delayed choice experiment;
we can choose to place detector $D^{(1)}$ at some $t_c\lesssim t_{int}$. 
In case of negative detection, an interaction at a distance between events 
$(x_2, t_{int})$, where the particle is localised (we could place a detector $D^{(2)}$ at $x_2$), and $(x_1, t_{int})$ for the detector would contradict RL. Another physical system must follow  path $1$ towards $x_1$, locally interacting at 
$(x_1,t_{int})$ with $D^{(1)}$ and giving way to the change of state of the detector. If both ARP and RL are fulfilled a particle is a composite of  a corpuscular subsystem, which follows path $2$, and a  distributed field\cite{dBro} following both paths. Suppression  of the 
$\sta{x_1}$ component of the field at $(x_1,t_{int})$,  
or modification of its state with a non destructive measurement or  some other interaction, 
determines the change of state of the composite system at $(x_f,t_f)$. $\sta{x_2}$ is a statistical representation of the subsystem made of particle (for negative detection at $x_1$) and $x_2$--field component, while $\sta{x_1}$ represents the $x_1$--field component. QM (and EQM) determines statistical distributions for outputs of measurement/detection of the corpuscular component; existence of an accompanying  de Broglie field does not contradict it.


 \section{\label{sec2}Extended phase--Hilbert spaces}

\subsection{Representation of physical magnitudes}

In Classical Mechanics\cite{AM}, a physical magnitude $\cal A$ is re\-pre\-sen\-ted by a function $x_A$ in the phase space of the system. $x_A$ can belong to a complete system of coordinates 
$\{x_A,\ldots\}$, whose values $\{a=x_A(s),\ldots\}$ characterize  states $s$ of the system. 
Trajectories are described by functions $x_A(s(t))$, $x_A$ having kinematic purposes. In Hamiltonian Mechanics, magnitude $\cal A$ has a second, dynamical representation:
the operator (vector field) $X_A=\{\cdot ,x_A \}$, where $\{,\}$ is the Poisson bracket.
The distinction be\-tween both representations is subtle, $\cal A$ and $\cal B$
seem to play similar roles in $\{x_B,x_A\}$. But the Poisson bracket has exclusively dynamical purposes, determining the law of evolution for a given Hamiltonian $H$

\[
\frac {dx_B}{dt}=\{x_B,H\}=\sum \frac {\partial H}{\partial x_A}\{x_B,x_A\}=
\sum c_AX_A(x_B) .
\]
In $\{x_B,x_A\}$, $x_B$  plays a kinematic role, $dx_B/dt$ describes an infinitesimal evolution $dx_B$ of magnitude $\cal B$ along a $dt$ time interval. On the other hand, the role of $x_A$, better $X_A$, is dynamical, it determines how much an $\cal A$ dependence of the Hamiltonian magnitude $\cal H$ adds up, with some intensity coefficient $c_A$, to the dynamical law of evolution  in the linear combination $d/dt=\sum c_AX_A$. 

In Quantum Mechanics, physical magnitudes $\cal A$ are elements of a non commutative algebra
$({\cal M},[,])$ ful\-fi\-lling canonical commutation rules. A maximal subalgebra 
${\cal F}=\{{\cal A},\ldots\}$ of commuting magnitudes determines ele\-men\-ta\-ry states labelled by specific values $(a,\ldots)$ of the magnitudes in $\cal F$. These states, understood as vectors $\sta{a,\ldots}$, generate through superposition a Hilbert space 
$H_{\cal F}^{^{\scriptscriptstyle {\scriptscriptstyle (\!Q\!M\!)}}}$, where $\cal A$ is represented by a self adjoint operator $A$ with eigenvalues $a$ and eigenstates 
$\sta{a,\ldots}$. Another ma\-xi\-mal abelian subalgebra 
${\cal F}'=\{{\cal B},\ldots\}$ defines the Hilbert space 
$H_{{\cal F}'}^{^{\scriptscriptstyle {\scriptscriptstyle (\!Q\!M\!)}}}$. Maps between bases 
$\sta{a,\ldots}$ and $\sta{b,\ldots}$, consistent with the commutation rules in 
$({\cal M},[,])$, allow to identify $H_{\cal F}$ with  $H_{{\cal F}'}$

\[
\xymatrix{H_{\cal F}^{^{\scriptscriptstyle {\scriptscriptstyle (\!Q\!M\!)}}}   
             \ar@<0.5ex>[rr] & & \ar@<0.5ex>[ll]  
             H_{{\cal F}'}^{^{\scriptscriptstyle {\scriptscriptstyle (\!Q\!M\!)}}} }         
\]
as well as to obtain the action of $B$ on 
$H_{\cal F}^{^{\scriptscriptstyle {\scriptscriptstyle (\!Q\!M\!)}}}$. The representations of physical magnitudes as labels of state and as dynamical operators are
{\bf  mixed up}, not properly distinguished: the number of labels $a$, coordinates for kinematic purposes, is unnecessarily constrained by the dimension of maximal abelian subalgebras of  $({\cal M},[,])$ with regards to the, conceptually different, dynamical structure $[,]$.  In this incomplete phase space the ARP is not automatically fulfilled.

\subsection{Extended Hilbert spaces}

The ARP is restated in an extended Hilbert space where kinematic and dynamical representations of magnitudes are distinguished\cite{mio2}. Elementary states 
$\sta{a,b,\ldots}$, labelled by 
$(a,b,\ldots)$, joint values of commuting and non commuting magnitudes in $\cal M$, generate 
$H_{\cal M}^{^{\scriptscriptstyle {\scriptscriptstyle (\!E\!Q\!M\!)}}}$;  $A$, $B$, \ldots, are linear operators in 
$ H_{\cal M}^{^{\scriptscriptstyle {\scriptscriptstyle (\!E\!Q\!M\!)}}}$, representation of 
$({\cal M},[,])$. 
$\sta{a,b,\ldots}$ is not eigenstate of $A$.
Standard QM Hilbert spaces 
$H_{\cal F}^{^{\scriptscriptstyle {\scriptscriptstyle (\!Q\!M\!)}}}$ are quotients

\[
\xymatrix{ & H_{\cal M}^{^{\scriptscriptstyle {\scriptscriptstyle (\!E\!Q\!M\!)}}} \ar_{\pi({\cal F})}[dl] \ar@<0.5ex>^{\pi({\cal F}')}[dr] & \\
             H_{\cal F}^{^{\scriptscriptstyle {\scriptscriptstyle (\!Q\!M\!)}}}   
              & &  
             H_{{\cal F}'}^{^{\scriptscriptstyle {\scriptscriptstyle (\!Q\!M\!)}}} }   
\]
defined by the natural projections $\pi({\cal F}): $
$H_{\cal M}^{^{\scriptscriptstyle {\scriptscriptstyle (\!E\!Q\!M\!)}}}$
$\to H_{\cal F}^{^{\scriptscriptstyle {\scriptscriptstyle (\!Q\!M\!)}}}$

\[
\sum_{(a,b,\cdots)\in {\cal M}} Z(a,b,\ldots)\sta{a,b,\ldots} \to
\]

\[ 
\sum_{(a,\cdots)\in {\cal F}} \left( \sum _{(b,\cdots)\in {\cal M}/{\cal F}}Z(a,b,\ldots)
\right) 
\sta{a,\ldots}
\] 

The QM distributions of amplitude, coefficients of the projected vectors, are  marginals of the EQM distribution, 
$Z(a,\ldots)=$ $\sum _{(b,\cdots)\in{\cal M}/{\cal F}}Z(a,b,\ldots)$. 
Subspace $H_{(a,\cdots)}$
$\subset H_{\cal M}^{^{\scriptscriptstyle {\scriptscriptstyle (\!E\!Q\!M\!)}}}$, made of all vectors with common labels $(a,\ldots)\in {\cal F}$,  
projects onto the ray  $\lambda \sta{a,\ldots}$ 
in $H_{\cal F}^{^{\scriptscriptstyle {\scriptscriptstyle (\!Q\!M\!)}}}$.  While in QM there is trivial action of $A$ over $\sta{a,\ldots}$, it 
becomes non trivial in $H_{(a,\cdots)}$, and the ARP
is automatically fulfilled. Mappings identifying different 
$H_{\cal F}^{^{\scriptscriptstyle {\scriptscriptstyle (\!Q\!M\!)}}}$ quotient spaces are superfluous: all standard projected vectors of state are obtained from a common vector in EQM,
and correspondingly all ob\-ser\-va\-ble distributions of probability are determined,
through marginals and Born's rule, from the common distribution of amplitude of the EQM state.

Many dynamical properties of a quantum system, as e.g. the quantised energy levels for a given Hamiltonian or the uncertainty relations,  are dictated by the algebraic structure of 
$({\cal M},[,])$. Therefore, these properties are common to all representations of the algebra in arbitrary Hilbert spaces. The following examples illustrate the main characteristics of the EQM formalism, the way it represents interference, the contextual character
of quantum distributions of probability, entanglement, etc.

\subsection{The two slit experiment}

For a spinless point particle, a ``universal covering'' phase space is the (ill defined) set of virtual paths in the path integral formalism\cite{Feynman}, each path with fixed amplitude  of probability 
$exp(iS[{\rm path}]/\hbar )$, where $S[{\rm path}]$ is the action integral. We can formally define  the state $\sta{S}=$ $\sum exp(iS[{\rm path}]/\hbar )$ $\sta{{\rm path}}$,  sum  along all paths allowed by the physical context.
In the  position representation the wave function
$\Psi (q) =$ $\sum  exp(iS[{\rm path(q)}]/\hbar )$ (${\rm path(q)}$ all paths with  endpoint $q$) is the marginal distribution of amplitude for the projection of $\sta{S}$ over the standard Hilbert space generated by $\{\sta{q}\}$. Similarly, the marginal 
$\sum  exp(iS[{\rm path(p)}]/\hbar )$ for fixed final momentum determines $\xi (p)$. 
 
On the opposite side of the largest phase space of paths, the paradigmatic two slit experiment is an exam\-ple of minimally extended phase space, with just an additional bivalued slit variable $s\in\{L,R\}$. $\Lambda(s,q)$, $q$ position at the final screen, is the marginal amplitude of all paths going from the source through slit $s$  to endpoint $q$. Paths from the source to points of the first screen out of the slits do not belong to the physical context of interest.  $\Psi(q)=\Lambda(L,q)+\Lambda(R,q)$ is the marginal corresponding to the projection $\sta{s,q}\to \sta{q}$, and $|\Psi(q)|^2$ describes the diffraction pattern. 
An alternative projection over the two dimensional  Hilbert space $\{\sta{s}\}$, restricting amplitudes (their phases) to integrals along paths with end points at the slits,
 determines $P(L)=P(R)=1/2$. It is applied to measurement of the slit variable.
In case of measurement, $\sta{s}$  evolves to $\Lambda'(s,q)$ with associated distribution of probability $|\Lambda'(s,q)|^2$, and no diffraction pattern is observed. Quantum probabilities are contextual, they depend on the superposition of amplitudes interfering in a given context. We can locate a particle detector, e.g. a narrow light beam, along the $R$ slit in such a way that the particle is unperturbed when going through the $L$ slit. Recording coincidence events between detection at $R$ and  the final screen we get the 
conditional distribution $P(L,q)$ for undetected particles. Without  local interaction between particle and detector the final state, statistical pattern, has changed. According to the ARP, some reaction on the light beam should happen. Perhaps using a coherent light beam 
some phase shift could be observed. According to RL, a de Broglie field component through the $R$ 
slit locally interacts with the light beam and arrives perturbed to the final screen.
If the interaction generates a stochastic shift of phase, the individual diffraction pattern at the final screen is also shifted. One  spot, final position of the corpuscular subsystem is re\-gis\-tered at each run; the overall statistical distribution is superposition of individual spots, each one associated to a randomly shifted diffraction pattern.

\subsection{Spinless point particle}

In the position representation for a spinless point particle, magnitudes $\cal Q$ and $\cal P$ are represented as $Q=q$, $P=-i\hbar\partial_q$ acting over the Hilbert space of square integrable complex functions of the spatial coordinates 
$\Psi(q)$. Alternatively, $Q=i\hbar\partial_p$ and $P=p$ are representations of the same Heisenberg algebra on the Hilbert space of square integrable functions of momentum. These Hilbert spaces are identified through the Fourier transform; in the language of path integral they are distinct quotient spaces of the set of virtual paths, defined by partitions into paths with common final position or momentum.
The superposition of amplitudes for all final momenta $p$ interfering at $q$ determines 
$\Psi(q)$. 
In EQM we can consider an ``incomplete'' projection, partition of the set of virtual paths
into sets with common final position and momentum, with marginal amplitudes some square integrable distribution $\Lambda (q,p)$ in the classical phase space. This defines a Hilbert space of elementary states  $\sta{q,p}$ with 
definite joint values of position and momentum. There is a formal, unobservable distribution of probability $|\Lambda (q,p)|^2$.  It is not Wigner's quasi probability 
distribution\cite{Wigner}  $W(q,p)$. $|\Psi (q)|^2$ is not the marginal of 
$|\Lambda (x,p)|^2$, because of the typical crossed interference terms in the  superposition of amplitudes. 
In \cite{torres} $Q_{\alpha}=\alpha q+i\hbar\partial_p$ 
and $P_{\beta}=\beta p-i\hbar\partial_q$, with $\alpha+\beta=1$, are the  representation of position and momentum dynamical operators. $\sta{q,p}$ is not eigenstate of the dynamical operator $Q$. The role of eigenstates is played here by  $Q$--invariant subspaces 
$H_{q_0}=\{\sta{q_0,p}\}$, which project onto  the rays  $\lambda \sta{q_0}$  in the standard  position representation. The action of $Q$ on $H_{q_0}$ is not trivial, it generates  evolution and allows to restate the ARP.
Obviously, there are more  possibilities of evolution
for $\Lambda(q,p,t)$ than for $\Psi(q,t)$.
Fourier projections $\Lambda(q,p) \to \Psi(q)$ $=\int \!dp K_{\alpha}(q,p) \Lambda(q,p)$, with $\alpha$--dependent Fourier kernels 
$K_{\alpha}(q,p)$\cite{torres} determined by the selection of bases, 
establish the correspondence with the standard formalism, and
all observable predictions of QM are reproduced. 
The observable distribution of probability $|\Psi (q)|^2$ associated to the marginal 
$\Psi(q)=\sum _p\Lambda(q,p)$
is a statistical representation of the superposition, interference between all $p$-components of the field at 
$q$.

\subsection{The singlet state}

In QM, when two magnitudes $A$ and $B$ commute the eigenvalues of a functionally dependent  $C=f(A,B)$  fulfil the same relation $c_{ij}=f(a_i,b_j)$, and $C$ is redundant. This is not the case for non commuting magnitudes. For example,  eigenvalues of the spin $1/2$ operators $S_{\phi}=$ $\cos \phi S_x + \sin \phi S_y$ are $s_{\phi}=$
 $s_x=s_y=\pm 1/2$ (in natural units), so that 
 $s_{\phi}$ $\neq s_x \cos \phi + s_y \sin \phi$.
Contrarily to a classical phase space, in EQM if $[A,B]\neq 0$, values $c$ of a functionally dependent magnitude $C=f(A,B)$ are independent,  not redundant labels of elementary states $\sta{a,b,c}$.
An elementary spin state in EQM  contains the values of spin in all directions of space,
a state $\sta{\sigma}$ cha\-rac\-te\-ri\-sed by a skew symmetric map $\sigma :S^2\to \{+,-\}$, 
$S^2$ the unit sphere. 
This is a ``universal covering'' phase space for spin variables,
analogous to the space of virtual paths for a spinless point particle. We can project  
$\sta{\sigma}$  onto a Hilbert space of spin in a finite  number of directions 
$\{{\bf n}_1, {\bf n}_2, \ldots, {\bf n}_N\}$, with spin coordinates 
$(s_1=\sigma ({\bf n}_1), s_2=\sigma ({\bf n}_2), \ldots)$ and elementary states
$\sta{s_1,s_2,\ldots,s_N}$. Two dimensional standard  Hilbert spaces are defined for one direction ${\bf n}$;
projections $\sta{\sigma} \to$ $\sta{\sigma ({\bf n})}$
define the basis $\{\sta{+},\sta{-}\}$ of spin up and down in direction ${\bf n}$. Changes of bases consistent with the commutation rules identify  Hilbert spaces for two directions. In EQM, each standard Hilbert space is a different quotient; the co\-rres\-pon\-dence between all projected states comes from a common state in the extended Hilbert space.

Spin coordinate $s_j$ in direction ${\bf n}_j$ has attached an elementary amplitude\cite{mio} 
$s_j{\bf N}_j$, where  ${\bf N}_j$ is the unit  quaternion  
$({\bf n}_{j}\cdot{\bf i}){\bf I} +$ $({\bf n}_{j}\cdot{\bf j}){\bf J} +$ 
$({\bf n}_{j}\cdot{\bf k}){\bf K}$,
$\bf I$, $\bf J$, $\bf K$ (and $\bf 1$)  generators of the quaternion numbers. $s_j{\bf N}_j$ plays the role of the unit complex amplitude $exp(iS/\hbar)$ in path integral. 
The total amplitude for state $\sta{s_1,s_2,\ldots,s_N}$
is the sum  $Z(s_1,s_2,\ldots,s_N)=$ $\sum s_j{\bf N}_j$. 
An orthodox state, e.g. spin up in direction $z$ (${\bf n}_1={\bf k}$),
is represented by the superposition of all  states $\sta{+_z,s_2,\ldots,s_N}$
with coefficients their corresponding amplitudes

\[
\sta{+_z} = \sum _{s_2,\ldots,s_N}({\bf K}+
\sum _{j\geq 2} s_j{\bf N}_j)\sta{+_z,s_2,\ldots,s_N}
\]
Projections over standard Hilbert spaces determine the right distributions of probability. $Z(+_z)= 2^{N-1}{\bf K}$, $Z(-_z)={\bf 0}$ 
gives $P(+_z)=1$, $P(-_z)=0$; 
from $Z(s_2)=$ $2^{N-2}({\bf K}+s_2{\bf N}_2)$
we get $P(s_2)=(1+s_2{\bf k}\cdot{\bf n}_2)/2$. Projected 
states, even of a normalised one, are generically not normalised.

If we interpret distributions of amplitude $Z(s_1,s_2,\ldots,s_N)$ as statistical, ensemble representations for the spin magnitudes of composite particle/field systems, a process in which the system splits into spatially divergent paths (e.g., through a Stern--Gerlach apparatus) corresponds to a splitting of ``spin field'' components (up and down),
while the particle follows one path. $\sta{+_z}$ represents  
the suppression of all $Z(-_z,s_2,\ldots,s_N)$ field components, when no physical context or experimental set up  rejoins these
field components previously separated.
The individual system has now definite $s^*_z=+$ value, hidden $s^*_j$ values and  both $s_j=\pm$
spin field components, for all $j\geq 2$. 

An isotropic physical process  generates 
particles, one at time, with unknown spin values in all directions. There is not a vector in QM representing this physical state, every vector in the Hilbert space is eigenstate of a spin operator in some direction. In EQM the vector

\[
\sta{S_0} = \sum _{s_1,s_2,\ldots,s_N}Z(s_1,s_2,\ldots,s_N)\sta{s_1,s_2,\ldots,s_N}
\]
represents the former physical state. Its projection over an arbitrary  standard Hilbert space determines the probabilities $P(s_j)=P(-s_j)=1/2$. EQM contains more states than standard QM. 
In $\sta{S_0}$ all spin field components are present and can interfere, while in a classical statistical ensemble of $\sta{+_z}$ and $\sta{-_z}$ there are not $\sta{+_z}$ field components in those systems of the ensemble represented by state $\sta{-_z}$.

A physical process generates pairs of correlated particles $\alpha$ and $\beta$ in the singlet spin state.
In EQM we can, in analogy with QM, assign to the composite the entangled vector
$\sta{S^{\alpha \beta}}=$

\[
\sum Z(s_1,s_2,\ldots,s_N)\sta{s_1,\ldots,s_N}^{\alpha}
\otimes \sta{-s_1,\ldots,-s_N}^{\beta}
\]
where the perfect (anti)correlation for outputs $s^{*\alpha}_j+s^{*\beta}_j=0$
is encoded, as in standard QM, in the coupling between vectors of each particle Hilbert space with opposite spin variables. Amplitudes outside the anti--diagonal  vanish. Alternatively, we can assign to the singlet state the factorized vector 
$\sta{S^{singlet}}=$ $\sta{S^{\alpha}_0}$ $\otimes^{corr}$ $\sta{S^{\beta}_0}$, 
and represent the correlated pairs through the functional relation between hidden values of spin  $s^{*\alpha}_j+s^{*\beta}_j=0$. The statistical description $\sta{S^{\alpha}_0}$ of particle $\alpha$ 
is independent of any measurement performed on  $\beta$. The correlation between Alice and Bob outputs of measurement $s^{*\alpha}_i$ and $s^{*\beta}_j$ on simultaneously generated particles, in directions ${\bf n}_i$ and ${\bf n}_j$ respectively, is equivalent to self correlation, specific of particle $\alpha$, between measured $s^{*\alpha}_i$ and hidden $s^{*\alpha}_j$; 
the output  $s^{*\beta}_j$ is used to infer the (hidden, counterfactual) $s^{*\alpha}_j$.

We can, through successive projections of $\sta{S_0}$, calculate the marginal amplitudes
$Z(s_i,s_j,s_k)$ (for a third direction ${\bf n}_k$), $Z(s_i,s_j)$
and $Z(s_i)$, from which we get $P(s_i,s_j,s_k)$, $P(s_i,s_j)$ and $P(s_i)$.
Only $P(s_i)$ is ob\-ser\-va\-ble because the  spin operators do not commute.
For generic magnitudes and states $P(s_i)$ will not match the marginal $\sum _jP(s_i,s_j)$ because of the interference terms in the sum of amplitudes. But $P(s_i)=\sum _jP(s_i,s_j)$ whenever both magnitudes are observable, either because 
the operators commute or because one of them can be inferred from 
measurement in a correlated companion.  State  $\sta{S^{\alpha}_0}$ of particle $\alpha$ fulfils the marginal probability condition for spin magnitudes in two arbitrary directions, independently of the existence of a correlated companion 
$\beta$. Measurement on $\beta$ allows to infer, without in any way perturbing the state of 
$\alpha$, a hidden value of spin for $\alpha$.
On the other hand, $\sum _kP(s_i,s_j,s_k) $ $\neq P(s_i,s_j)$, i.e.,
in\-ter\-fe\-rence terms in the sum of amplitudes $\sum _kZ(s_i,s_j,s_k)$
do not vanish. There is no way, for particles in state 
$\sta{S^{singlet}}$, to measure/infer three values of spin for $\alpha$.
The phenomenon is analogous to the two slit experiment: spin field components in any third
direction $s_k=\pm$ interfere as both slits wave components, determining
$|Z(s_i,s_j,+_k)+Z(s_i,s_j,-_k)|^2$ $=$ 
$|Z(s_i,s_j,+_k)|^2+$ $|Z(s_i,s_j,-_k)|^2$ $+{\rm interf}$. Because of the non vanishing interference terms $P(s_i,s_j)$ does not match $P(s_i,s_j,+_k)$ $+P(s_i,s_j,+_k)$, in analogy with $P(q)\neq$ $P(R,q)+P(L,q)$ because  $|\Psi(L,q)+\Psi(R,q)|^2=$
$|\Psi(L,q)|^2+|\Psi(R,q)|^2$ $+{\rm interf}$.

\section{Quantum statistical theory of field/particle composites}

Measurement is an interaction, and according to the ARP it causes a reaction on the measured system.  Let us consider, for academic purposes,
a classical statistical theory for microscopic systems in which the unavoidable reaction under measurement can not be neglected.
A physical process or preparation procedure generates one system at a time.
There is, for the ensemble, an unknown  distribution of probability $P(a_i,b_j)$ for two non commuting magnitudes $\cal A$ and $\cal B$. If we perform measurements of $\cal A$ in a statistical sample determining $P(a_i)$, followed by measurements of $\cal B$ in the output states,
the observed $P(a_i,b^{out}_j)$ will differ from
the unknown initial $P(a_i,b_j)$; we can not apply, by hypothesis, the approximation
$b^{out}_j=b_j+\bigtriangleup b_j\simeq b_j$. Using the observable 
$P(a_i)$ and $P(b_j)$ (obtained from another statistical sample) we can look for a joint probability distribution fulfilling the linear equations

\begin{equation}
P(a_i) \!=\! \sum _j P(a_i,b_j),\,\,\,  P(b_j) \!=\! \sum _i P(a_i,b_j),
\end{equation}
and inequalities $P(a_i,b_j)\!\geq\! 0$. Each measured system is in a state with definite (unknown) values of both magnitudes, so that there is solution, not necessarily unique. Alternatively, when the system under study is jointly generated with a correlated companion, in such a way that output of measurement of a magnitude ${\cal B}^{corr}$ determines $b=f(b^{corr})$, the observable $P(a_i,b^{corr}_j)$ determines  $P(a_i,b_j=f(b^{corr}_j))$.

QM is a statistical theory, but not the former classical statistical theory for microscopic systems. QM distributions of probability are contextual, which means that  there is not 
necessarily a joint $P(a_i,b_j)$ for non commuting magnitudes $A$ and $B$ and arbitrary state $\sta{S}$. The system (1)
is not generically compatible in QM, with 
$P(a_i)=|\stacc{a_i}\sta{S}|^2$,
$P(b_j)=|\stacc{b_j}\sta{S}|^2$ associated to the QM vector of state $\sta{S}$. 
If we ignore the positivity condition, the remaining system of linear equations becomes compatible. For two non commuting magnitudes $A$ and $B$, and a unit vector of state 
$\sta{S}$, a (not necessarily unique) solution is

\[
W(a,b)=\frac{1}{2}\left(\stacc{S}\sta{a}\stacc{a}\sta{b}\stacc{b}\sta{S} + {\rm cc}\right)
\]
${\rm cc}$ the complex conjugate. It can be easily checked that 
$W(q,p)$ is Wigner's quasi probability distribution when 
$\stacc{q}\sta{S}=\Psi(q)$ and $\stacc{p}\sta{S}=\xi(p)$ are position and momentum representations for a spinless point particle. For $N$ magnitudes a solution is

\vskip -0.2truecm

\[
W(a,b,\ldots,c) = \frac{1}{N!}(\stacc{S}\sta{a}\stacc{a}\sta{b}
\cdots \stacc{c}\sta{S} + \cdots)
\]
including all permutations of $(a,b,\ldots,c)$. Anyhow, quasi probabilities are obtained from the more fundamental distributions of amplitude.

If  a quantum  particle
is not isolated, if it is accompanied by a de Broglie field,
the attempt to describe an ensemble of field/particle composites through a distribution of probability $P(a_i,b_j)$ in the phase space of corpuscular variables would ignore the field degrees of freedom,  interference phenomena and the interaction between particle and field subsystems. 
The system (1) is trivially compatible whenever $P(a_i,b_j)$ is observable. In particular, it is compatible for commuting magnitudes; the unavoidable reaction to measurement does not, in an ideal case, modify the value of the commuting variable.
If a non commuting magnitude $B$ can be inferred from measurement of $B^{corr}$
in a correlated companion,  
$[A,B^{corr}]=0$ and $P(a_i,b_j)$ is observable. Alice knows, since the space time measurement event $\alpha$ of $A$, the output of an hypothetical Bob's measurement, at space time  event $\beta$,
of the correlated magnitude $A^{corr}$, no matter if $A^{corr}$ has already  been measured at  some $\beta$ in the past of $\alpha$, or it will  be measured in the future of $\alpha$, or there is spatial separation between $\alpha$ and the hypothetical $\beta$ measurement events. QM distributions of probability are contextual because there is not a $P(a_i,b_j)$ for non commuting magnitudes 
in arbitrary states. In EQM, 
a $P_{EQM}(a_i,b_j)$ is obtained applying Born's rule to 
$\sta{S}= \sum z_{ij}\sta{a_i,b_j}$, $P_{EQM}(a_i,b_j)=|z_{ij}|^2$ for $\sta{S}$ normalized. Ge\-ne\-ri\-ca\-lly,  marginals of $P_{EQM}$ do not match  $P(a_i)$ and $P(b_j)$ because of non vanishing interference terms. The consistency conditions for observability of $P_{EQM}$ are 

\[
P(a_i)\equiv\frac {1}{{\cal N}_A}|\sum _j z_{ij}|^2 = \sum _jP_{EQM}(a_i,b_j)
=\sum _j|z_{ij}|^2
\]

\[
P(b_j)\equiv\frac {1}{{\cal N}_B}|\sum _i z_{ij}|^2 = \sum _iP_{EQM}(a_i,b_j)
=\sum _i|z_{ij}|^2
\]
${\cal N}_{A}$,  ${\cal N}_{B}$ normalization factors.
QM and EQM states, distributions of amplitude of probability, represent an ensemble of 
composite systems made of corpuscular and wave like subsystems. Neither QM nor EQM 
contain a mathematical representation of an individual composite system, its evolution laws and interactions between the physical field and the particle. The amplitudes and re\-la\-ti\-ve phases of the complex (or quaternion) distribution encode, in the statistical ensemble representation,  superposition and interference phenomena between different field components and the particle/field interaction. An accompanying de Broglie field can give account of the contextual character of quantum probabilities, as in the paradigmatic two slit experiment.

An elementary particle, composite of field and corpuscular subsystems,
interacts  with an external system. If the external system is macroscopic
its state is not completely known.
The interaction causes statistical decoherence between field components,
different unknown phase shifts at each repetition of the process.
Statistical decoherence gives account of  the projection rule under measurement.
For example, in the two slit experiment measurement of the slit variable
generates relative phase shift between both slit wave components,
and correspondingly spatial shift of the diffraction pattern. Statistical superposition 
of randomly shifted diffraction patterns is equivalent to
suppression of the interference behaviour.
Only a few macroscopic systems show long range coherence: superconductivity, Bose--Einstein condensates,  coherent light beams, \ldots
The classical limit of QM is approached when superposition and interference phenomena of the field component can be neglected; 
in a macroscopic composite of many particles only additive (average) corpuscular degrees of freedom remain.

\end{document}